\documentstyle[12pt]{article}

\renewcommand{\thefootnote}{\fnsymbol{footnote}}
\newcommand{\EQ}{\begin{equation}}
\newcommand{\EN}{\end{equation}}
\newcommand{\bea}{\begin{eqnarray}}
\newcommand{\ena}{\end{eqnarray}}
\newcommand{\vs}[1]{\vspace{#1 mm}}
\newcommand{\hs}[1]{\hspace{#1 mm}}

\renewcommand{\b}{\beta}

\newcommand{\pa}{\partial}

\newcommand{\uda}{\nearrow \kern-1em \searrow}

\begin{document}

\topmargin 0pt
\oddsidemargin 5mm

\begin{titlepage}
\setcounter{page}{0}
\begin{flushright}
OU-HET 215 \\
June, 1995
\end{flushright}

\vs{12}
\begin{center}
{\Large Composite Neutrinos and Double Beta Decay}
\vs{15}

{\large
Eiichi Takasugi\footnote{e-mail address: takasugi@phys.wani.osaka-u.ac.jp}}\\
\vs{8}
{\em Department of Physics, \\
Osaka University \\ Toyonaka, Osaka 560, Japan} \\
\end{center}
\vs{6}

\centerline{{\bf{Abstract}}}

Neutrinoless double beta decay $(\b\b)_{0\nu}$ occurs through the magnetic
coupling of dimension five, $\lambda_W^{(\nu*)}/m_{\nu*}$,  among the excited
electron neutrino $\nu^*$, electron and $W$ boson if $\nu^*$  is a massive
Majorana neutrino. If the coupling  is not small, i.e., $\lambda_W^{(\nu*)}>1$
 the mass of the excited neutrino must not be gless than the  $Z$ boson mass,
$m_Z$.  Since $\nu^*$ contributes in the  $(\b\b)_{0\nu}$  decay  as a vertual
state,  this decay will give an oppotunity to explore the much heavier mass
region of  $\nu^*$.

In this paper, we present the decay formula of $(\b\b)_{0\nu}$ decay through
the $\nu^*$ exchange and discuss the constraint on the coupling constant and
the mass of the excited neutrino. By comparing the recent data for ${}^{76}$Ge,
we find $\lambda_W^{(\nu*)}({1\rm TeV}/m_{\nu*})) (m_N/{1\rm TeV})^{\frac 12}<
4.1\cdot 10^{-3}$ where    $m_N$ is the Majorana mass of the excited electron
neutrino.
If $m_N=m_{\nu*}$ and $\lambda_W^{(\nu*)}>1$, we find the mass bound for the
excited  Majorana neutrino as $m_{\nu^*} > 5.9\cdot 10^4$TeV.

In order to obtain the constraint on the composite scale $\Lambda$, we have to
specify the model.  For  the mirror type and the homodoublet type models,
$\lambda_W^{(\nu*)}/m_{\nu*}=f/(\sqrt 2 \Lambda)$ where $f$ is the relative
strength of gauge couplings. Then, we obtain  $\Lambda > 170 f (m_N/{1\rm
TeV})^{\frac 12}$TeV.  For the sequential type  model,
$\lambda/m_{\nu*}=fv/(\sqrt 2 \Lambda^2)$ where $v$ is the vacuum expectation
value of the dopublet Higgs boson, i.e., $v=$250GeV. In this model, we find
$\Lambda > 6.6 f^{\frac 12} (m_N/{1\rm TeV})^{\frac 14}$TeV.

\end{titlepage}
\newpage
\renewcommand{\thefootnote}{\arabic{footnote}}
\setcounter{footnote}{0}
\section{Introduction}

If neutrinos are composite particles, there exist  the excited neutrinos
which couple to the ground state leptons by the dimension five magnetic
coupling\cite{CMS},\cite{HKZ}.  This interaction is expressed as\cite{PDG}
\bea
L_{int}=g\frac{\lambda_W^{(\nu^*)} }{ m_{\nu^*}}\bar e \sigma^{\mu\nu}(\eta_L^*
R+\eta_R^* L)\nu^* \pa_\mu W_\nu^-  + h.c. ,
\ena
where $\nu^*$ is a heavy excited electron neutrino, $L=(1-\gamma_5)/2$,
$R=(1+\gamma_5)/2$, $m_{\nu*}$ is the mass demension which is of order the mass
of $\nu^*$. This interaction is derived by the $SU(2)\times U(1)$gauge
invariant form  and parameters $\eta_L$ and  $\eta_R$ should satisfy the
following constraints by the chirarity conservation\cite{PDG},
\bea
\mid \eta_L \mid ^2 + \mid \eta_R \mid^2 =1, \hs{3} \eta_L \eta_R=0.
\ena
The extensive search of $\nu^*$ have been made by many groups\cite{MG} and
found that $m_{\nu*} > 91$GeV by assuming that  $\lambda_Z > 1$ which is the
coupling for $\nu^* \to e Z$ decay similarly defined  to
$\lambda_W^{(\nu^*)}$. So far, the severe mass bound in the region  $m_{\nu*} >
m_Z$ is not obtained.

The purpose of this paper is to explore the mass renge $m_{\nu*} > m_Z$ by
using the neutrinoless double beta decay $(\b\b)_{0\nu}$  by assuming that
$\nu^*$ is  a massive Majorana neutrino.  Then the $(\b\b)_{0\nu}$ decay
occurs through $\nu^*$  exchange.   Since $\nu^*$  enters as a virtual state,
we can investigate heavy $\nu^*$.  Panella and Srivastava\cite{PS} was the
first to investigate $(\b\b)_{0\nu}$ decay, but unfortunately their formula is
incorrect. Here we shall present the correct expression for the decay .

We start from the interaction in Eq.(1) which leads to the effective four point
interaction between leptons and hadrons as
\bea
L_{eff}=-G_{eff}\bar e \sigma^{\mu\nu}(\eta_L^* R+
\eta_R^* L)\nu^* \pa_\mu J_\nu^\dagger+ h.c. ,
\ena
where $J_\nu^\dagger$ is the hadronic current and
\bea
G_{eff}=2G_F \frac{g\lambda_W^{(\nu^*)} }{m_{\nu^*}}.
\ena
I
\section{Decay formula of neutrinoless double beta decay}

In the second order perturbation of the effective interaction in Eq.(3), the
$(\b\b)_{0\nu}$ decay  takes place and the S-matrix for this decay is given by
\bea
S=i{{G_{eff}^2} \over{2 (2\pi)^4}} \int dxdy\int dq{{m_N e^{-iq(x-y)}} \over
{q^2-m^2_N}}\bar e(x) \sigma^{\mu\nu}\sigma^{\rho\sigma}
(\eta_L^{*2 }R+\eta_R^{*2} L)e^C(y) \pa_\mu^x\pa_\rho^yT(J_\nu^\dagger(x)
J_\sigma^\dagger(y)) ,
\ena
where $e^C$ is the charge conjugation of $e$, i.e., $e^C=C\bar e^T$. In the
follwoing, we take  the S-wave of electron wave function which is given by
\bea
<0\mid e(x) \mid p>= \psi_S(\epsilon) e^{-i\epsilon x^0};
\hs{3}
\psi_S(\epsilon)=\sqrt{{\epsilon+m} \over {2\epsilon}}\left(\begin{array}{c}
\chi_s \\
\frac{\vec \sigma\cdot \vec p}{\epsilon+m} \chi_s
\end{array} \right)
F_0(Z,\epsilon) ,
\ena
where $\epsilon$ is the energy of electron and $F_0(Z,\epsilon) $ is the
relativistic Coulomb factor defined in Eq.(3.1.25) in Ref.*. The S-wave
function  is independent of the space coordinate. With this wave function,  we
obtain
\bea
S_{fi}&=&i{{G_{eff}^2} \over{2(2\pi)^4}} \int dxdy\int dq{{e^{-iq(x-y)}m_N}
\over {q^2-m^2_N}}<N_f\mid T(J_\nu^\dagger(x)J_\sigma^\dagger(y)\mid N_i>\times
\nonumber \\
&& \left[t^{\nu\sigma}(\epsilon_1,\epsilon_2,q^0,\vec q)e^{i(\epsilon_2
x^0+\epsilon_1 y^0)}
-(\epsilon_1 \leftrightarrow \epsilon_2)\right]
\ena
where
\bea
t^{\nu\sigma}(\epsilon_1,\epsilon_2,q^0,\vec q)=\bar \psi_S(\epsilon_2)
\sigma^{\mu\nu}\sigma^{\rho\sigma}
(\eta_L^{*2 }R+\eta_R^{*2} L)\psi_S^C(\epsilon_1)(q_\mu-\epsilon_2  g_{\mu 0})
(q_\rho+\epsilon_1 g_{\rho 0})
\ena

Now  we perform the $q^0$ integration first and then do the $x^0$ and
$y^0$ integration. We found

\bea
R_{fi}&=&{{G_{eff}^2} \over{\sqrt{2!} 2(2\pi)^3}} \int d\vec x d\vec y\int
d\vec q\frac{m_N}{\omega}e^{i\vec q\cdot (\vec x-\vec y)}\times \nonumber \\
&&\sum_n \left[ \frac{<N_f\mid J_\nu^\dagger(\vec x)\mid n><n\mid
J_\sigma^\dagger(\vec y) \mid
N_i>}{\omega+E_n-E_i+\epsilon_1}t^{\nu\sigma}(\epsilon_1,\epsilon_2,\omega,\vec
q)
-(\epsilon_1 \leftrightarrow \epsilon_2)\right] ,
\ena
where $\sqrt{2!}$  in the denominator is the statistical factor.
In this expression, $q^0=\omega=\sqrt{m_N^2+q^2}$ with $q=\mid \vec q \mid$
which comes from the pole at $q^0=\omega$ of the neutrino propergator. The
contribution from the
$q^0=-\omega$ pole turns out to be equal to the one from $q^0=\omega$ pole in
the approximation to keep S-wave of electron wave function.
\par
So far our formulae is exact up to the S-wave approximation of electron
wave function. Next, we make following approximations: (1) We take the closure
approximation where $E_n$ is replaced by the average value
$<E_n>$ so that the sum of the intermediate states can be taken.
(2)$\epsilon_1$ and $\epsilon_2$   in the numerator is neglected,  $(q_\mu
-\epsilon_2  g_{\mu 0})(q_\rho+\epsilon_1 g_{\rho 0}) \cong q_\mu q_\rho$.  (3)
The energy denominators
$\omega+<E_n>-E_i+\epsilon_1=\omega+<E_n>-(\frac{E_i+E_f}{2})+(\frac{\epsilon_1-\epsilon_2}{2})$  is replaced by
$\mu_0 m_e\equiv  <E_n>-\frac{(M_i+M_f)}{2}$.  Here we replaced $E_i$ and $E_f$
by their masses $M_i$ and $M_f$. The approximation (2) and (3) are  valid
because  $\epsilon_1$ and $\epsilon_2$ are of order 1MeV  which is much smaller
than $\omega=\sqrt{m_N^2+q^2}$, the  average value of $q$  which is of order
$1/R$   with $R$ beeing the nuclear radius and
$\mu_0 m_e$.

With these approximations, we get
\bea
R_{fi}&=&{{G_{eff}^2} \over{\sqrt{2!} 2(2\pi)^3}} \int d\vec x d\vec y\int
d\vec q\frac{m_N}{\omega}e^{i\vec q\cdot (\vec x-\vec y)} \frac{<N_f\mid
J_\nu^\dagger(\vec x) J_\sigma^\dagger(\vec y) \mid N_i>}{\omega+\mu_0 m_e}
\times\nonumber \\
&&q_\mu q_\rho \bar \psi_S(\epsilon_2)
\{\sigma^{\mu\nu},\sigma^{\rho\sigma}\}(\eta_L^{*2 }R+\eta_R^{*2}
L)\psi_S^C(\epsilon_1) .
\ena

Now we use the identity $\{\sigma^{\mu\nu},\sigma^{\rho\sigma}\} =
2(g^{\mu\rho}
g^{\nu\sigma}-g^{\mu\sigma}g^{\nu\rho}-i\epsilon^{\mu\nu\rho\sigma}\gamma_5)$
and  the non-relativistic approximation of the hadronic current as
\bea
J_\mu^\dagger(x)=\sum_n J_\mu^\dagger (n) \delta (\vec x-\vec r_n),\hs{2}
J_\mu^\dagger (n) =\tau_n^+ (g_V g_{\mu 0}+g_A \delta_{\mu j}
\sigma_n^j)F(q^2),
\ena
where $r_n$ is the position of the n-th nucleon in the nucleus and  $F(q^2)$ is
the form factor defined by
\bea
F(q^2)=(\frac{1}{1+(q^2/m_A^2)})^2,
\ena
with the value $m_A=0.85$MeV.  We obtain
\bea
R_{fi}&=&{{G_{eff}^2} \over{\sqrt{2!}(2\pi)^3}} \sum_{n,m} \int d\vec q\frac{
e^{i\vec q\cdot r_{nm}}}{\omega(\omega+\mu_0 m)} <N_f\mid J_\nu^\dagger (n )
J_\sigma^\dagger(m) \mid N_i>\times \nonumber \\
&&m_N(m_N^2 g^{\nu\sigma}-q^\nu q^\sigma)\bar \psi_S(\epsilon_2)
(\eta_L^{*2 }R+\eta_R^{*2} L)\psi_S^C(\epsilon_1),
\ena
where $r_{nm}=\vec r_n-\vec r_m$.
It should be noted that $m_N^2 g^{\nu\sigma}$, $q^0q^0=\omega^2$  and $q^jq^l$
terms  contribute only  to $0^+ \to 0^+$ transition, while $q^0 q^j=\omega q^j$
term
contributes to $0^+ \to 0^-$ transition.  The $0^+ \to 2^+$ transition does
not occur because we keep the S-wave of electron wave function.
\par
In the following, we concentrate on the $0^+ \to 0^+$ transition.  Sinc we
consider a heavy composite neutrino $m_N $ is much greater than $m_A$,
we expand $\omega$ and $\omega +\mu_0m_e$ in the $q$ integration in the power
of $1/m_N$. We find
\bea
\frac{m_N(m_N^2 g^{\nu\sigma}-q^\nu q^\sigma)}{\omega(\omega+\mu_0
m_e)}&&J_\nu^\dagger(n)J_\sigma^\dagger(m )
\cong-g_A^2 \tau_n^+\tau_m^+ \{(m_N-\mu_0 m_e+\frac{(\mu_0 m_e)^2}{m_N}
)\vec\sigma_n\cdot\vec\sigma_m \nonumber \\
&&+[(\frac{g_V}{g_A})^2-\vec\sigma_n\cdot\vec\sigma_m]\frac{q^2}{m_N}
+\sigma_n^j\sigma_m^k\frac{q^jq^k}{m_N}\} F^2(q^2).
\ena
Thus we obtain

\bea
R_{fi}&=&{{(G_{eff} m_A g_A)^2} \over{\sqrt{2!} 4\pi R}} \psi_S(\epsilon_2)
(\eta_L^{*2 }R+\eta_R^{*2} L)\psi_S^C(\epsilon_1)\nonumber \\
&&\{(m_N -\mu_0 m_e+\frac{(\mu_0 m_e)^2}{m_N} )  M_{GT,N} +\frac{m_A^2}{m_N}(
(\frac{g_V}{g_A})^2 M_F'-\frac 23     M_{GT}'-\frac 13 M_T')\}.
\ena
where
\bea
M_{GT,N}&=&<N_f\mid\sum_{n\ne m}\tau_n^{(+)}\tau_m^{(+)} \vec
\sigma_n\cdot\vec\sigma_m (\frac {R}{r_{nm}})F_N(x_A)\mid N_i>,\\
M_{F}'&=&<N_f\mid\sum_{n\ne m}\tau_n^{(+)}\tau_m^{(+)} (\frac
{R}{r_{nm}})F_4(x_A)\mid N_i>,\\
M_{GT}'&=&<N_f\mid\sum_{n\ne m}\tau_n^{(+)}\tau_m^{(+)} \vec
\sigma_n\cdot\vec\sigma_m (\frac {R}{r_{nm}})F_4(x_A)\mid N_i>,
\ena
\bea
M_T'=<N_f \mid\sum_{n\ne m}\tau_n^{(+)}\tau_m^{(+)} \{ 3(\vec \sigma_n\cdot\vec
r_{nm})  (\vec \sigma_m\cdot\vec r_{nm}) - \vec\sigma_n\cdot\vec\sigma_m
\}(\frac {R}{r_{nm}})F_5(x_A)\mid N_i>,
\ena
where $x_A=m_Ar_{nm}$ and neutrino potentials are
\bea
F_N(x)&=&\frac{x}{48}(3+3x+x^2)e^{-x},\hs{2}
F_4(x)=\frac{x}{48}(3+3x-x^2)e^{-x},\nonumber \\
&&F_5(x)=\frac{x^3}{48}e^{-x}
\ena
\par
After taking the spin sum and performing the phase-space integration, we
find the half-life of the  transition of the neutrinoless double beta decay due
to the heavy composite neutrino as
\bea
T^{-1}(0^+ \to 0^+)&=&4\left( \frac{\lambda m_A}{m_{\nu*}}\right)^4
\left(\frac{m_A}{m_e}\right)^2 G_{01}
\mid (\frac{m_N}{m_A}-\frac{\mu_0 m_e}{m_A}+\frac{(\mu_0
m_e)^2}{m_Nm_A})M_{GT,N} \nonumber \\
&&+\frac{m_A}{m_N}[\frac{g_V}{g_A})^2 M_F'-\frac 23 M_{GT}'-\frac 13 M_T']\mid
^2 ,
\ena
where $G_{01}$ is the phase space factor defined in Eq.(3.5.17a) in Ref.6.
 The above expression  is obtained by taking $\mid
\eta_L\mid^4+\mid\eta_R\mid^4=1$ which is valid for both chirality cases
because we imposed the chirality conservation in Eq.(2). That is, we consider
only two cases,   $(\eta_L,\eta_R)=(1,0)$ or $(0,1)$.

\section{Constraint from neutrinoless double beta decay}

In the following, we analyse the constraint on the coupling by using the
experimental half-life limit of  ${}^{76}{\rm Ge}$ which is measured by the
Heidelberg-Moscow collaboration\cite{HM},
\bea
T(0^+ \to 0^+ :{}^{76}{\rm Ge}) > 5.6\cdot10^{24}yr \hs{3} 90\% c.l. .
\ena
We derive the constraint on composite parameters from this data. By using the
values of nuclear matrix elements obtained by Hirsch, Klapdor-Kleingrothaus and
Kovalenko[7],
\bea
M_{GT,N}=0.113, \hs{1} M_{F}'=3.06\cdot10^{-3}, \hs{1}
M_{GT}'=-7.70\cdot10^{-3}, \hs{1} M_T'=-3.09\cdot10^{-3},
\ena
and the phase space factor $G_{01}=6.4\cdot10^{-15}/yr$ given in Ref.6,
we find the half-life of the $0^+ \to 0^+$ transition for ${}^{76}{\rm Ge}$ is
given by
\bea
T^{-1}=\left( \frac{\lambda m_A}{m_{\nu*}}\right)^4
\mid (\frac{m_N}{m_A}-\frac{\mu_0 m_e}{m_A}+\frac{(\mu_0 m_e)^2}{m_Nm_A}+
7.2\cdot 10^{-2}\frac{m_A}{m_N}\mid ^2  9.1\cdot10^{-10} /yr .
\ena
By comparing this formula to the data, we find by taking,
\bea
\left( \frac{\lambda m_A}{m_{\nu*}}\right)^2
\mid (\frac{m_N}{m_A}-\frac{\mu_0 m_e}{m_A}+\frac{(\mu_0 m_e)^2}{M_Nm_A}+
7.2\cdot 10^{-2}\frac{m_A}{m_N}\mid < 1.4\cdot 10^{-8}.
\ena

Our formula in Eq.(21) is valif for the case where the excited neutrino is a
Majorana neutrino which is much heavier than  $m_A=0.85$GeV. In this case, the
dominant term in Eq.(25) is the one which is proportional to the excited
neutrino mass $m_N$ because $\mu_0 m_e$ is of order 10MeV. Thus the constraint
becomes a simple form as
\bea
\lambda\left(\frac{1{\rm TeV}}{m_{\nu*}}\right) \cdot \left(\frac{m_N}{{\rm
1TeV}}\right)^{\frac 12}<4.1\cdot 10^{-3}.
\ena

\section{Limit on the excited neutrino mass and the composite scale}

Firstly, we shall discuss about the excited neutrino mass. If we assume that
$m_N=m_{\nu*}$, we obtain
\bea
m_{\nu*} > 5.9\cdot 10^4 {\rm TeV} \hs{3} (\lambda_W^{\nu*} > 1).
\ena
This is the most stringent  bound so far.

Secondly we discuss the bound on the composite scale.  Since the relation
between the coupling constant and the composite scale is model dependent,  we
have to confine   some  specific models\cite{PDG}.

\begin{enumerate}
\item Sequential type model

We consider the sequential type model where the excited leptons are assigned as

\bea
L_L^*=\left(\begin{array}{c}\nu^*\\ e^*\end{array}\right)_L, \hs{2}
[\nu^*_R],\hs{2} e^*_R ,
\ena
where $[\nu_R^*]$ indicates that there are two cases, with and without
$\nu_R^*$.

There are two kinds of  gauge invariant interactions. If $\nu_R^*$ exists, the
interaction  is given by

\bea
L_{int}=\frac{ 1}{2\Lambda}\bar \ell_L
\sigma^{\mu\nu}(gf\frac{\tau^a}{2}W_{\mu\nu}^a +g'f'\frac{Y}{2 }B_{\mu\nu})\phi
\nu_R^* + h.c.  ,
\ena
where $\ell_L$ is the left-handed ground state lepton doublet and $\phi$ is the
doublet Higgs boson. This interaction  gives the one  in Eq.(1) with
\bea
\frac{\lambda_W^{(\nu^*)} }{ m_{\nu^*}}=\frac{fv}{\sqrt 2 \Lambda^2}
  \hs{3}{\rm with} \hs{3}(\eta_L, \eta_R)=(1,0),
\ena
where $v$ is the vacuum expectation value of $\sqrt 2 \phi$ which is about
250GeV and $\nu_R^*$ mediates the decay.

If $\nu_R^*$ does not exist, the  other type interaction will be present
\bea
L_{int}=\frac{ 1}{2\Lambda}\bar e_R \phi^\dagger
\sigma^{\mu\nu}(gf\frac{\tau^a}{2}W_{\mu\nu}^a +g'f'\frac{Y}{2 }B_{\mu\nu})
L_L^* + h.c.  .
\ena
This interaction gives
\bea
\frac{\lambda_W^{(\nu^*)} }{ m_{\nu^*}}=\frac{fv}{\sqrt 2 \Lambda^2}
  \hs{3}{\rm with} \hs{3}(\eta_L, \eta_R)=(0,1)
\ena
and $\nu_L^*$ meadiates.

For both cases, we find
\bea
\Lambda > 6.6f^{\frac 12} (\frac{m_N}{1\rm TeV})^{\frac 14}{\rm TeV}.
\ena

\item  Mirror type

In this model, the multiplet of the excited leptons are
\bea
[\nu^*_L] ,\hs{2} e^*_L , \hs{2} , L_R^*= \left(\begin{array}{c}\nu^*\\
e^*\end{array}\right)_R .
\ena

The gauge invariant interaction is
\bea
L_{int}=\frac{ 1}{2\Lambda}\bar \ell_L
\sigma^{\mu\nu}(gf\frac{\tau^a}{2}W_{\mu\nu}^a +g'f'\frac{Y}{2
}B_{\mu\nu})L_R^* + h.c.  ,
\ena
which leads the coupling
\bea
\frac{\lambda_W^{(\nu^*)} }{ m_{\nu^*}}=\frac{f}{\sqrt 2 \Lambda}
  \hs{3}{\rm with} \hs{3} (\eta_L,\eta_R)=(1,0)
\ena
and $\nu_R^*$ mediates.
In this case, we find
\bea
\Lambda > 170 f (\frac{m_N}{1\rm TeV})^{\frac 12}{\rm TeV}.
\ena

\item  Homodoublet type

In this model, the multiplet of  excited leptons are
\bea
L_L^*=\left(\begin{array}{c}\nu^*\\ e^*\end{array}\right)_L, \hs{2}
L_R^*=\left(\begin{array}{c}\nu^*\\ e^*\end{array}\right)_R .
\ena
In this case, the gauge invariant interaction is the same as Mirror case so
that we have the same coupling in Eq.(*) and $\nu_R^*$ mediates. Thus, we
obtain

\bea
\Lambda > 170 f (\frac{m_N}{1\rm TeV})^{\frac 12}{\rm TeV}.
\ena

\end{enumerate}

In the above, we obtain rather strong constraints on the mass of $\nu^*$ and
the composite scale. These constraints may be modified if there is a mixing for
excited neutrinos. There are two types of mixing; the flavor mixing and the
$\nu_L^* - \nu_R^*$ mixing. If such mixings are present, the  excited  neutrino
mass $m_N$ should be replaced by a effective mass as
\bea
m_N \hs{2}\Rightarrow \hs{2}<m_{\nu*}> \equiv \sum_j (U_{ej}^2\hs{2}{\rm or}
\hs{2}V_{ej}^2)m_j ,
\ena
as the one appeared in the neutrino mass contribution to the ordinary
neutrinoless double beta decay. The mixing parameters are defined by
\bea
\nu_L^*= \sum_j U_{ej}N_{jL}^* ,\hs{3}\nu_R^*= \sum_j V_{ej}N_{jR}^* ,
\ena
where $N_j^*$ is the mass-eigenstate Majorana neutrino with the mass $m_j$.
The effective mass $<m_{\nu*}> $  in essence comes from the Majorana mass of $
\nu_L^*$ or $\nu_R^*$. Thus, if the main part of $\nu^*$ mass comes from the
Dirac mass, then the constraint is not valid. In particular, if $\nu^*$ is a
Dirac particle, there is no constraint from the neutrinoless double beta decay.

\vs{5}
Acknowledgment

The author would like to thank to O. Panella,  K. Higashijima and K. Hagiwara
for useful discussions and comments.


\newpage
\begin{thebibliography}{99}
\bibitem{CMS}
N. Cabbibo, L. Maiani and Y. Srinivastava,
Phys. Lett.  139B, 459 (1984).

\bibitem{ HKZ}
K. Hagiwara, S. Komamiya and D. Zeppenfeld,
Z. Phys. C29, 115 (1985).

\bibitem{PDG}
Particle Data Group,
Review of Particle Properties, Phys. Rev. D50, 1173 (1994).

\bibitem{MG}
L3 Collaboration,
Phys. Rep. C236, 1 (1993);

ALEPH Collaboration,
Phys. Rep. C216, 253 (1992);

See references in Particle Data Group.

\bibitem{PS}
O. Panella and Y.N. Srivastava,
College de Frace Preprint, LPC 94 39.



\bibitem{DKT}
M. Doi, T. Kotani and E. Takasugi,
Prog. Theor. Phys.  Supplement 83, 1 (1985).

\bibitem{HKK}
M. Hirsh, H.V. Klapdor-Kleingrothaus and S.G. Kovalenko,
Max Planck Institute preprint, MPI-H-V 6-1995.

\bibitem{HM}
HEIDERBERG-MOSCOW Collaboration: A. Balysh et al.,
Proceedings of the 27th Int. Conf. on Heigh Energy Physics, 20th-27th July
1994, Glasgow.


\end{thebibliography}
\end{document}